\documentclass[a4paper,fleqn,usenatbib]{mnras}
\usepackage[T1]{fontenc}
\usepackage{ae,aecompl}

\usepackage{graphicx}   
\usepackage{amsmath}    
\usepackage{amssymb}    

\newcommand{\figureref}[1]{Fig.~\ref{#1}}

\newcommand{\tableref}[1]{Table~\ref{#1}}

\title[DIG in disc galaxies with CR feedback]{Radiative transfer calculations of
the diffuse ionised gas in disc galaxies with cosmic ray feedback.}

\author[B. Vandenbroucke et al.]{
Bert Vandenbroucke,$^{1}$\thanks{E-mail~: bv7@st-andrews.ac.uk}
Kenneth Wood,$^{1}$
Philipp Girichidis,$^{2}$
Alex Hill,$^{3,4,5}$
\newauthor
Thomas Peters$^{6}$
\\
\\
$^1$SUPA, School of Physics \& Astronomy, University of St Andrews, North Haugh,
St Andrews, KY16 9SS, United Kingdom
\\
$^2$Leibniz-Institut f\"{u}r Astrophysik Potsdam (AIP), An der Sternwarte 16,
14482 Potsdam, Germany
\\
$^3$Department of Physics and Astronomy, University of British Columbia,
Vancouver, British Columbia V6T 1Z1, Canada
\\
$^4$Space Science Institute, Boulder, CO USA
\\
$^5$National Research Council Canada, Herzberg Program in Astronomy and
Astrophysics, Dominion Radio Astrophysical Observatory,\\ PO Box 248, Penticton,
British Columbia V2A 6J9, Canada
\\
$^6$Max-Planck-Institut f\"{u}r Astrophysik, Karl-Schwarzschild-Str. 1, D-85748
Garching, Germany
}

\date{Accepted XXX. Received YYY; in original form ZZZ}

\pubyear{2017}

\begin{document}

\label{firstpage}
\pagerange{\pageref{firstpage}--\pageref{lastpage}}
\maketitle

\begin{abstract}
The large vertical scale heights of the diffuse ionised gas (DIG) in disc
galaxies are challenging to model, as hydrodynamical models including only
thermal feedback seem to be unable to support gas at these heights. In this
paper, we use a three dimensional Monte Carlo radiation transfer code to
post-process disc simulations of the Simulating the Life-Cycle of Molecular
Clouds (SILCC) project that include feedback by cosmic rays. We show that the
more extended discs in simulations including cosmic ray feedback naturally lead
to larger scale heights for the DIG which are more in line with observed scale
heights. We also show that including a fiducial cosmic ray heating
term in our model can help to increase the temperature as a function of disc
scale height, but fails to reproduce observed DIG nitrogen and sulphur forbidden
line intensities. We show that, to reproduce these line emissions, we require a
heating mechanism that affects gas over a larger density range than is achieved
by cosmic ray heating, which can be achieved by fine tuning the total
luminosity of ionising sources to get an apropriate ionising spectrum as a
function of scale height. This result sheds a new light on the relation between
forbidden line emissions and temperature profiles for realistic DIG gas
distributions.
\end{abstract}


\section{Introduction}

Observations of the extended ISM in the Milky Way and other galaxies have shown
the existence of a diffuse ionised component of the ISM with scale heights of
more than a kpc above the star forming disc \citep{2009Haffner}. Due to its
temperatures of $\sim{}10^4~{\rm{}K}$, this component is usually referred to as
the warm ionised medium (WIM). The optical emission line ratios and inferred
temperature ($\sim{}10^4~{\rm{}K}$) of the gas imply that it is photoionised, so
that it is also referred to as the diffuse ionised gas (DIG). The most obvious
possible source of ionising radiation is the O and B stars in the galactic disc
itself \citep{1990Reynolds}. Previous studies have shown that this radiation is
sufficient to ionise the DIG at high altitudes, provided that the gas density
distribution is not smooth \citep{2005Wood, 2010Wood, 2014Barnes}. A fractal or
lognormal density distribution is implied by observations and arises naturally
from turbulence in the DIG \citep{1997Elmegreen, 2008Hill, 2008Berkhuijsen}. The
importance of photoionisation indicates that the gas is heated by UV radiation.

To reproduce the observed scale height of the DIG in 3D models of the galactic
disc, a detailed model of the structure of the disc is needed.
\citet{2014Barnes} used the models of \citet{2012Hill}, which include thermal
stellar feedback in a full magnetohydrodynamical (MHD) treatment of the ISM, and
showed that they fail to reproduce the required densities at high altitude that
could explain the observed scale height of the DIG. Furthermore, they also show
the need for an additional source of heating that would explain the observed
intensities of the [N\textsc{ii}] $6584$~\AA{} and [S\textsc{ii}] $6725$~\AA{}
forbidden lines as a function of the H$\alpha{}$ intensity. These results
indicate that a simple model including only thermal supernova feedback is unable
to support an extended DIG and unable to heat the gas to high enough
temperatures. This means we either need a more complete model of stellar
feedback that also includes radiative feedback and stellar winds that can change
the structure of the ISM \citep{2017Gatto, 2017Peters}, an additional from of
heating that affects gas at low densities \citep{1999Reynolds}, or both.

To address these issues, we repeat the analysis of \citet{2014Barnes} for a new
sample of MHD simulations that include cosmic ray feedback
\citep{2016Girichidisb}. Unlike thermal feedback, cosmic ray feedback does not
couple directly to the local gas but can diffuse to higher altitudes through the
magnetic field. This means it can heat the gas more efficiently and support a
thicker disc. It also provides a non local heating term that could provide the
additional heating needed to explain observed line ratios.

The structure of this paper is as follows. In Section 2, we summarize the new
ionisation code we used, and give some more details about the simulation models.
We then show that our models are converged (Section 3.1), and illustrate the
effect of changing various model parameters (Section 3.2). We then move on to a
full description of the time evolution of the models (Section 3.3), and of the
emission line maps that we produce (Section 3.4). We end with our conclusions.

\section{Method}

\subsection{Code}

We use the new Monte Carlo radiative transfer code \textsc{CMacIonize}
\citep{2017Vandenbroucke}
\footnote{\url{https://github.com/bwvdnbro/CMacIonize}}, which is essentially a
rewritten version of the ionisation code of \citet{2004Wood}. The code employs a
basic model whereby ionising radiation from one or more sources is propagated
through a discrete density grid, and only absorption and re-emission by hydrogen
and helium are taken into account. As \citet{2010Wood} showed, the effect of
dust absorption is minimal and can hence be neglected. The ionising part of the
source spectrum is sampled using a number of discrete photon packets, which are
emitted isotropically from the source location(s). Each photon packet is then
followed while it traverses the simulation volume, until a randomly sampled
optical depth is reached. At this point, the photon packet is absorbed, and is
re-emitted at a randomly sampled frequency. We do not adopt the so called on the
spot approximation, whereby re-emitted photons are absorbed locally, and instead
treat re-emitted photons in the same way as source photons. Photon packets whose
frequency drops below the ionisation treshold, or that leave the simulation box
without scattering (through a non periodic boundary), are removed from the
system.

For each cell in the simulation volume, we keep track of the path lengths
traversed by photon packets that pass through it. After a sufficient number of
photon packets has been evolved in this way, the path length counters in each
cell can be used to obtain a good approximation of the mean ionising intensities
in that cell. We then use these values to calculate the local ionisation
equilibrium. This will likely change the properties of the cell, so that we need
to repeat the whole process until convergence is reached.

As the ionisation equilibrium also depends on the temperature, we also need to
keep track of heating and cooling processes that might affect the thermal
equilibrium of the ISM. We consider heating caused by direct UV absorption by
hydrogen and helium, heating due to indirect, ``on the spot'' absorption of
He\textsc{i} Ly$\alpha{}$, and heating caused by absorption by polycyclic
aromatic hydrocarbons (PAHs). As an optional extra heating term, we also
consider heating by cosmic rays, following \citet{2013Wiener}. Apart from
cooling by recombination of ionised hydrogen and helium, and cooling by
free-free emission (bremsstrahlung), we also consider a number of metals to
obtain cooling rates~: C, N, O, Ne, and S. To this end, intensity counters are
also stored for different ions of these metals~: C$^+$, C$^{++}$. N$^0$, N$^+$,
N$^{++}$, O$^0$, O$^+$, O$^{++}$, Ne$^+$, Ne$^{++}$, S$^{+}$, S$^{++}$, and
S$^{+++}$.

Since the full combined ionisation and temperature algorithm is more
computationally expensive than a more approximate approach whereby the
temperature is kept fixed, we first explore the available parameter space using
a fixed temperature of 8,000~K (which corresponds to the average equilibrium
temperature of the observed DIG), and only use the combined algorithm to obtain
line intensities for the models with the most realistic parameter values.

In runs that use the full version of the code, we can use the resulting
ionisation structure of the various coolants and the equilibrium temperature to
compute forbidden line emission \citep{2004Wood, 2017Vandenbroucke}. This
allows us to produce line emission maps and line ratios that can be directly
compared with observations.

\subsection{ISM density field}

We use simulated density fields from the Simulating the Life-Cycle of Molecular
Clouds (SILCC) project \citep{2015Walch, 2016Girichidisa}, more specifically,
the three different models described by \citet{2016Girichidisb}~: a model with
only thermal stellar feedback, a model with only cosmic ray stellar feedback,
and a model with both forms of feedback. Thermal stellar feedback
consists of adding $10^{51}~{\rm{}erg}$ of energy to the gas surrounding a
supernova event. This happens either as pure energy injection in regions where
the Sedov-Taylor expansion of the supernova is resolved, or as a momentum
injection in regions were it is not resolved. Cosmic ray feedback consists of
adding $10^{50}~{\rm{}erg}$ of energy to the cosmic ray energy equation.
This equation is evolved as an extra separate equation during the
magnetohydrodynamical integration, and couples to the hydrodynamics as an extra
pressure term in the momentum and energy equations. The cosmic ray energy
equation assumes a simplified transport equation based on an isotropic particle
distribution function, and neglects the effect of cosmic ray streaming. None of
the models we post-process includes a prescription for photoionisation feedback.

The simulations themselves were run in a box of
$2\times{}2\times{}\pm{}20~{\rm{}kpc}$, with periodic $x$ and $y$ boundaries,
and using an adaptive mesh with a resolution of $15.6~{\rm{}pc}$
($128\times{}128\times{}2560$ cells) in the high resolution region. The three
simulations respectively cover an evolution over $257.3~{\rm{}Myr}$,
$263.7~{\rm{}Myr}$, and $256.9~{\rm{}Myr}$. In this work, we mainly focus on the
results after $250~{\rm{}Myr}$ of evolution, although we will also briefly
discuss the time evolution of the ionisation structure. It is worth noting that
the \citet{2016Girichidisb} simulations do not include a self-consistent
modeling of star formation using sink particles, nor early stellar feedback from
radiation or stellar winds, like e.g. \citet{2017Peters}.

To perform the post-processing with our radiative transfer code, we resample the
density grids on a Cartesian grid of $128\times{}128\times{}256$ cells, in a box
with dimensions $2\times{}2\times{}\pm{}2~{\rm{}kpc}$, and with the same
periodic boundaries as the simulations themselves. Extending the box vertically
to larger sizes does not change our results, since the density at large vertical
heights drops to negligible values. The adopted number of cells was found to
provide an optimal trade-off between accuracy and computational efficiency. We
tested the convergence of our results using a simulation with a higher
resolution, which oversamples the original SILCC data. The average vertical
column density of the gas within our box is $\sim{}10~{\rm{}M_\odot{}~pc}^{-2}$
($\sim{}10^{21}~{\rm{}atoms~cm}^{-2}$) for all three models at $250~{\rm{}Myr}$,
consistent with the average H\textsc{i} column density in the Milky Way
\citep{1990Dickey}, and in line with the column densities found close to Milky
Way spiral arms \citep{2003Nakanishi}.

The simulated density fields use a fixed solar metallictiy, so we have to make
some assumptions for the abundances of He and the coolants we track. We will use
the same values as in \citet{2014Barnes}~: ${\rm{}O/H} = 4.3 \times{} 10^{-4}$,
${\rm{}N/H} = 6.5 \times{} 10^{-5}$ \citep{2004Simpson,2009Jenkins}, ${\rm{}S/H}
= 1.4 \times{} 10^{-5}$ \citep{2009Daflon}, ${\rm{}He/H} = 0.1$, and
${\rm{}Ne/H} = 1.17 \times{} 10^{-4}$ \citep{2000Mathis}.

\subsection{Ionising sources}
The ionising sources are luminous O and B stars. \citet{1982Garmany} found an
average stellar surface density of 24 kpc$^{-2}$ for O stars in the solar
neighbourhood, so we randomly sample 96 sources within our
$2\times{}2~{\rm{}kpc}^2$ box. The $x$ and $y$ coordinates of the sources are
uniformly sampled, while the $z$ coordinate is sampled from a normal
distribution. The scale height of the stellar disc is a parameter for our
models. We use a value of $63~{\rm{}pc}$ \citep{2001MaizApellaniz} for most of
our models. The typical ionising luminosity of an O star is $\sim{}10^{49}$
s$^{-1}$ \citep{2003Sternberg}, however it is not a priori clear how much of
this radiation makes it out of the dense \textsc{Hii} region surrounding the
star and into the interstellar medium \citep{2010Wood}. We therefore treat the
fraction of the $\sim{}10^{49}$ s$^{-1}$ luminosity that is effectively used to
ionise the ISM as a second parameter of our source model.

The spectrum of the ionising sources is based on the stellar models of
\citet{2001Pauldrach}; we use the data tables compiled by \citet{2003Sternberg}
to obtain the spectrum of a 40,000~K star with a surface gravity of
$\log\left({g}/{\left({\rm{}cm~s^{-1}}\right)}\right) = 3.40$, on which we
linearly interpolate.

The density fields we use contain asymmetries that can potentially propagate
into our radiative transfer calculation, especially when our source distribution
is sampled randomly and independent of the density field. We address this issue
by testing the robustness of our results against a change in the seed for the
random generator used to sample our source distribution.

\subsection{Simulations}


In total, 52 radiative transfer simulations were run for each of the three
feedback models: 1 reference model, 10 models to test the convergence of our
model, 6 to test the effect of parameter changes in the ionising source model,
10 more advanced models used to calculate emission lines, and 25 models to show
the time evolution of the ionised disc. The latter used SILCC snapshots at
different times throughout the simulation, while all the other simulations were
run using the snapshot at $250~{\rm{}Myr}$. The different parameters and our
adopted naming convention for the first two groups of models are shown in
\tableref{table_naming}. Our reference model uses 20 iterations for a
$128\times{}128\times{}256$ cell grid with $10^7$ photon packets, with a source
scale height of $63~{\rm{}pc}$ and a per source luminosity of
$4.26\times{}10^{49}$~s$^{-1}$, and fixed source positions (set by the fixed
random seed 42). Note that simulations Ci20g128p7 and Ir42l49s063 both refer to
this reference model.

The more advanced models use the same parameters, but with $10^8$ photon
packets, to ensure converged coolant fractions.

\begin{table*}
\caption{Naming convention for the convergence tests and source model parameter
exploration.
\label{table_naming}}
\begin{tabular}{l c c c}
\hline
simulation name & number of iterations & grid resolution & number of photon
packets \\
\hline
Ci20g128p7 & 20 & $128\times{}128\times{}256$ & $10^7$ \\
\hline
Ci05g128p7 & 5 & $128\times{}128\times{}256$ & $10^7$ \\
Ci10g128p7 & 10 & $128\times{}128\times{}256$ & $10^7$ \\
Ci15g128p7 & 15 & $128\times{}128\times{}256$ & $10^7$ \\
Ci25g128p7 & 25 & $128\times{}128\times{}256$ & $10^7$ \\
\hline
Ci20g064p7 & 20 & $64\times{}64\times{}128$ & $10^7$ \\
Ci20g256p7 & 20 & $256\times{}256\times{}512$ & $10^7$ \\
Ci20g256p8 & 20 & $256\times{}256\times{}512$ & $10^8$ \\
\hline
Ci20g128p5 & 20 & $128\times{}128\times{}256$ & $10^5$ \\
Ci20g128p6 & 20 & $128\times{}128\times{}256$ & $10^6$ \\
Ci20g128p8 & 20 & $128\times{}128\times{}256$ & $10^8$ \\
\hline
\end{tabular}
\begin{tabular}{l c c c}
\hline
simulation name & random seed & ionising luminosity & scale height \\
 & & (${\rm{}s}^{-1}~{\rm{}source}^{-1}$) & (pc) \\
\hline
Ir42l49s063 & 42 & $4.26\times{}10^{49}$ & 63.0 \\
\hline
Ir19l49s063 & 19 & $4.26\times{}10^{49}$ & 63.0 \\
Ir55l49s063 & 55 & $4.26\times{}10^{49}$ & 63.0 \\
\hline
Ir42l48s063 & 42 & $4.26\times{}10^{48}$ & 63.0 \\
Ir42l50s063 & 42 & $4.26\times{}10^{50}$ & 63.0 \\
\hline
Ir42l49s032 & 42 & $4.26\times{}10^{49}$ & 31.5 \\
Ir42l49s126 & 42 & $4.26\times{}10^{49}$ & 126.0 \\
\hline
\end{tabular}
\end{table*}

\section{Results and discussion}

\subsection{Convergence}

\begin{figure*}
\centering{}
\includegraphics[width=\textwidth]{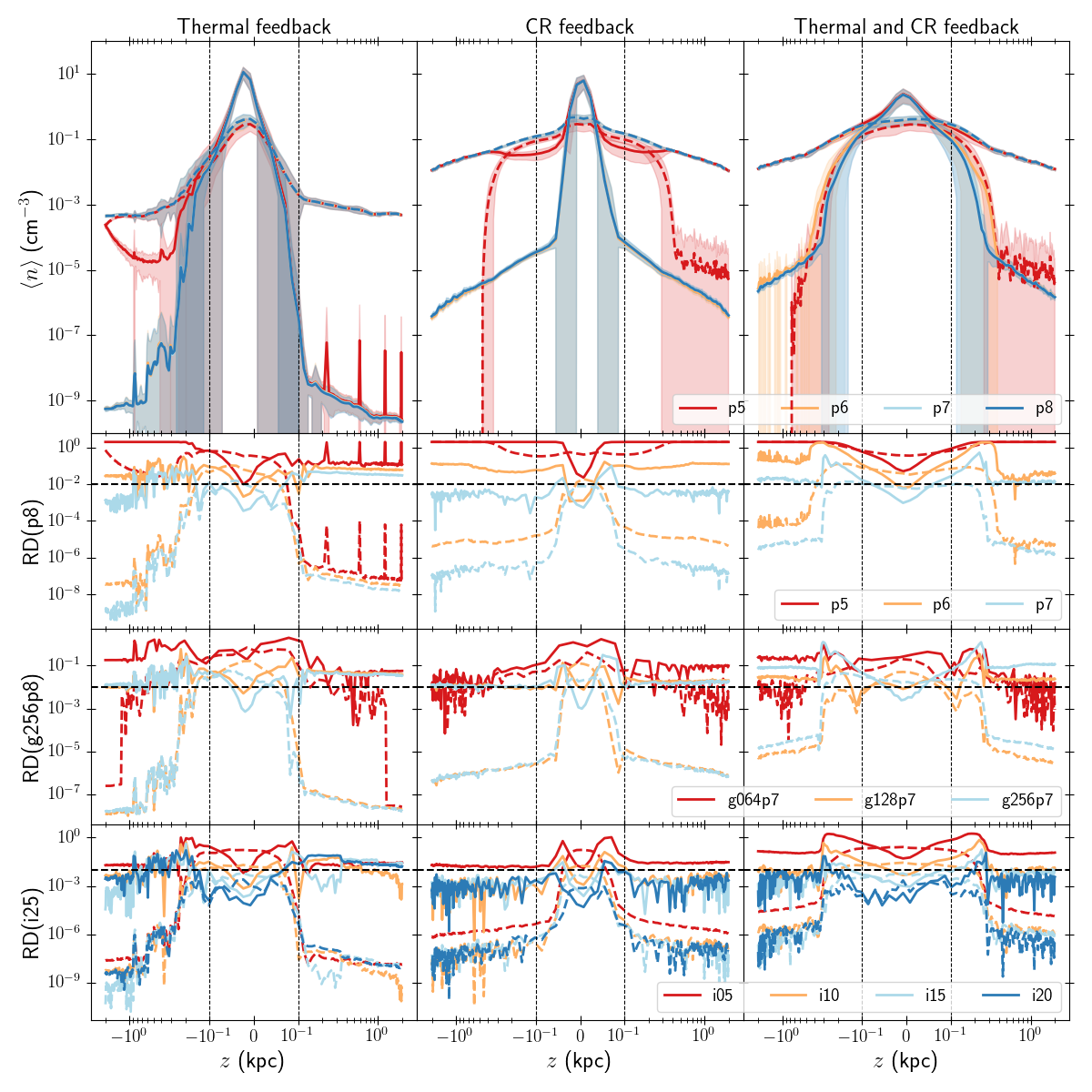}
\caption{\emph{Top row~:} average neutral (full line) and ionised (dotted line)
density for a given number of photon packets used in our algorithm. The shaded
regions represent the scatter within slices of equal height above the disc. For
clarity, the region close to the midplane has been plotted on a linear scale,
while the outer regions are plotted on a logarithmic scale. The dashed vertical
lines indicate the transition from the linear to the logarithmic region. Note
that the large scatter near the midplane causes some of the shaded regions to
have negative lower bounds, which can not be represented on the logarithmic
scale. \emph{Other rows~:} relative difference between the average densities of
the indicated models and the reference model for different model parameters:
\emph{top~:} number of photon packets, \emph{middle~:} grid resolution,
\emph{bottom~:} number of iterations. The dashed horizontal lines represent the
target 1~\% convergence limit.
\label{fig_convergence_everything}}
\end{figure*}

Before we can test convergence of our results, we have to make clear what
convergence means in our case. We are interested in the vertical scale of the
neutral and ionised disc, so we require vertical disc profiles which are
sufficiently converged. To quantify convergence, we compare profiles for the
average ionised and neutral gas densities. These are defined as respectively the
spatially averaged ionised and neutral gas densities in $2\times{}2$~kpc planar
slices with constant height $z$ above or below the disc. Examples of
density profiles are shown in the top panel of
\figureref{fig_convergence_everything}.

For each parameter of interest, we compare profiles for different
parameter values with a reference model by computing the relative difference
\begin{equation}
{\rm{}RD}(X,z) = 2 \left| \frac{n(z) - n_X(z)}{n(z) + n_X(z)} \right|,
\end{equation}
with $X$ the parameter value that identifies the reference model. A model is
considered sufficiently converged if the relative difference is of the level of
1~\% or less for all values of $z$.

We start with the number of iterations of our algorithm that is needed to obtain
converged densities. The bottom row of
\figureref{fig_convergence_everything} shows the relative difference between the
density profiles after 5, 10, 15 and 20 iterations, and the reference result
after 25 iterations. The largest differences between the low iteration number
models and the reference model are located near the central disc. The ionised
disc between 0.5 and 2~kpc is converged to 1~\% accuracy after 20 iterations,
which is the number we will use for all our simulations.

A very important parameter in our algorithm is the number of photon packets used
to discretize the stellar radiation field. If this number is too low,
discretization error will dominate our simulation, and the radiation field will
be unable to spread throughout the simulation box.
The top rows of \figureref{fig_convergence_everything} show the
average densities for different numbers of photon packets, and for a fixed grid
resolution of $128\times{}128\times{}256$ cells. It is immediately clear that
the results for the simulations with only $10^5$ photon packets are
qualitatively different from those of the other simulations, illustrating how
the radiation field is unable to efficiently ionise the box if too few photon
packets are used. Just as in the case of the number of iterations, using a low
number of photons also affects the scale height of the ionised disc. The
relative difference drops below 1~\% for the simulations with $10^7$ photon
packets, and this is the number that we will use in consecutive simulations.

A last important parameter for the convergence of our models is the grid
resolution. Unlike the number of iterations or photon packets, this parameter is
not immediately linked to our photoionisation algorithm. This means that we can
obtain a converged photoionisation solution for any given grid resolution by
using enough photon packets and iterations. If the grid samples the density
field very badly, then this solution will not be converged to the solution
imposed by the underlying density field, but from the point of view of the
photoionisation algorithm, it will be converged nonetheless. Conversely,
parameters that lead to a converged result for one grid resolution, will not
necessarily work for another resolution. The central row of 
\figureref{fig_convergence_everything} shows how using a
$256\times{}256\times{}512$ grid with only $10^7$ photon packets seems to change
our result, indicating that we need at least this resolution to obtain a
converged result. However, if we rerun the same model with 10 times more photon
packets, then the solution looks much more similar to the
$128\times{}128\times{}256$ result, showing that the latter resolution is
sufficient for convergence. This is to be expected, as this is the resolution
used by \citet{2016Girichidisb}, so we are effectively overresolving the input
density field with a $256\times{}256\times{}512$ grid. As pointed out by
\citet{2004deAvillez, 2012Hill}, this resolution does not guarantee converged
density fields in the midplane of the SILCC simulations, and might cause us to
overestimate the ionising luminosity necessary to ionise the extended disc.

We conclude that 20 iterations of our algorithm, using a grid of
$128\times{}128\times{}256$ cells and $10^7$ photon packets, is sufficient to
obtain converged vertical density profiles for the ionised disc.

\subsection{Ionising source model}

\begin{figure*}
\centering{}
\includegraphics[width=\textwidth]{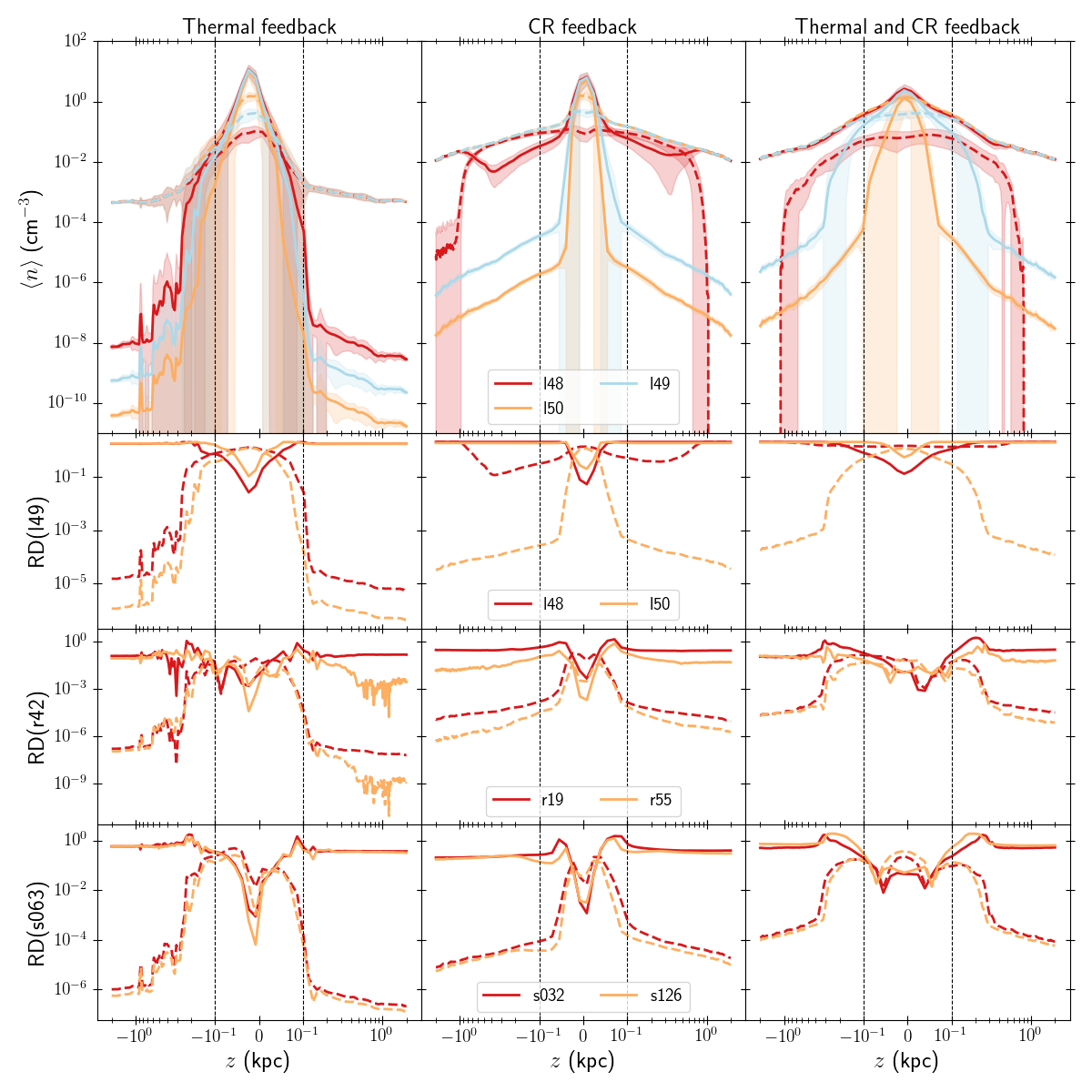}
\caption{\emph{Top row~:} average neutral (full line) and ionised (dotted line)
density for a given ionising luminosity per ionising source. The shaded regions
represent the scatter within slices of equal height above the disc. For clarity,
the region close to the midplane has been plotted on a linear scale, while the
outer regions are plotted on a logarithmic scale. The dashed vertical lines
indicate the transition from the linear to the logarithmic region. Note that the
large scatter near the midplane causes some of the shaded regions to have
negative lower bounds, which can not be represented on the logarithmic scale.
\emph{Bottom rows~:} relative difference between the average densities of
the indicated models and the reference model for different model parameters:
\emph{top~:} ionising luminosity per source, \emph{middle~:} random seed used to
generate source positions, \emph{bottom~:} scale height of the ionising disc.
\label{fig_ionising_sources_everything}}
\end{figure*}

Before we can discuss the effect of changing the source luminosity or scale
height of the source distribution on our results, we need to quantify the
scatter introduced by our sampling of the source distribution. To this end, we
rerun each of our models with two different seeds for the random generator that
generates the positions of our discrete sources.

The central row of \figureref{fig_ionising_sources_everything}
shows the relative difference in densities for models with different random
seeds (and hence different positions for the 96 ionising sources). Apart from a
small difference in the width of the central density peak, the relative
difference between the different models is of the order of 10\% in the neutral
gas density. There is a clear asymmetry in the neutral gas disc for the SILCC
model without cosmic ray feedback, which shows that the exact positions of the
sources can enhance asymmetries present in the density field. However, for the
ionised gas density, and especially the extended disc which is of most interest
to us, the relative differences between different source models are small. We
conclude that the overall structure of our solution is independent of the
numerical details of our ionising source model, so that we can use a single
realization of the source model to study the structure of the ionised disc.

We already mentioned above that there is some uncertainty on the ionising
luminosity of the individual sources in our model, since we do not know how much
radiation is absorbed in the local envelope surrounding each source, which we do
not resolve. To address this uncertainty, we run models with different ionising
luminosities per ionising source. The top rows of
\figureref{fig_ionising_sources_everything} show the density profiles for three
different values of the luminosity~: the generic value of $4.26\times{}10^{49}$
s$^{-1}$, a value that is 10 times lower, and a value that is 10 times higher.
The latter is not really physical, but it is instructive to see how it affects
our results.

It is immediately clear from the figure that a low value of the luminosity can
lead to results that are qualitatively completely different~: if the luminosity
is too low, then there is insufficient radiation to ionise the entire box, and
we end up with an extended neutral disc instead of an ionised disc. However,
when the luminosity is high enough, an ionised disc is created, the structure of
which is relatively independent of the total ionising luminosity. Just as with
the different random seeds above, we see that a different luminosity has some
effect on the width of the central density profile, but does not really affect
the extended disc (provided the luminosity is high enough to ionise it). The
neutral density profile is affected, but only in the regions where the neutral
gas fraction is already quite low.

The bottom row of \figureref{fig_ionising_sources_everything} shows
the relative differences for models with different values for the ionising
source distribution scale height~: the generic value of 63 pc, a value which is
half of that, and a value which is twice the generic value. It is clear that the
scale height of the sources has a small effect on the value of the neutral
fraction, as it is easier for hard ionizing radiation to escape from the dense
inner disc if the source distribution is more extended.

We conclude that the extended ionised disc in our models is robust against
changes in our ionising source model, provided that the ionising luminosity of
the combined sources is sufficient to ionise the disc. These results are in
agreement with previous results of \citet{2010Wood}.

\subsection{Time evolution}

\begin{figure*}
\centering{}
\includegraphics[width=\textwidth]{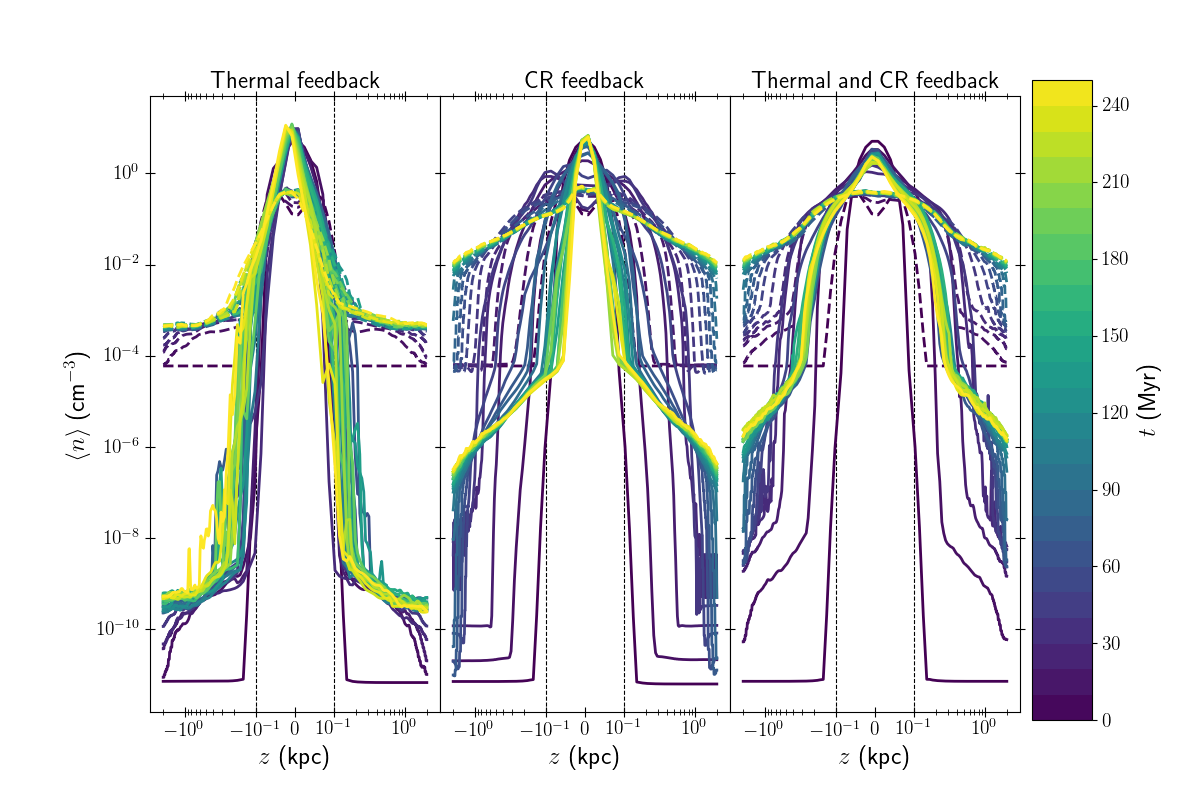}
\caption{Average neutral (full line) and ionised (dotted line) density profile
at different times during the simulations. For clarity, the region close to the
midplane has been plotted on a linear scale, while the outer regions are plotted
on a logarithmic scale. The dashed vertical lines indicate the transition from
the linear to the logarithmic region.
\label{fig_time_evolution}}
\end{figure*}

\figureref{fig_time_evolution} shows the build-up of the density profiles over
time, by post processing snapshots at a 10 Myr interval from time 0 Myr to the
final snapshot at 250 Myr we extensively discussed before, for each of the three
feedback models. For the model with only thermal feedback, the profiles quickly
reach a stable configuration in less than 50 Myr, for the models with cosmic ray
feedback, the build-up is slower and it takes approximately 100 Myr to reach a
stable disc.

The figure also shows how all three models at first have very low disc
densities, which slowly increase over time. For the model with only thermal
feedback, this increase is limited, while for the other two models the disc
growth is more extended.

It is important to note that these results were obtained in post processing, so
there is no dynamical effect of the ionizing radiation on the evolution of the
simulation model (other than the feedback model employed by the SILCC
simulations themselves). Furthermore, all our results used the same source
model, independent of the evolution of the underlying SILCC model.

\subsection{Emission lines}

\begin{figure}
\centering{}
\includegraphics[width=0.5\textwidth]{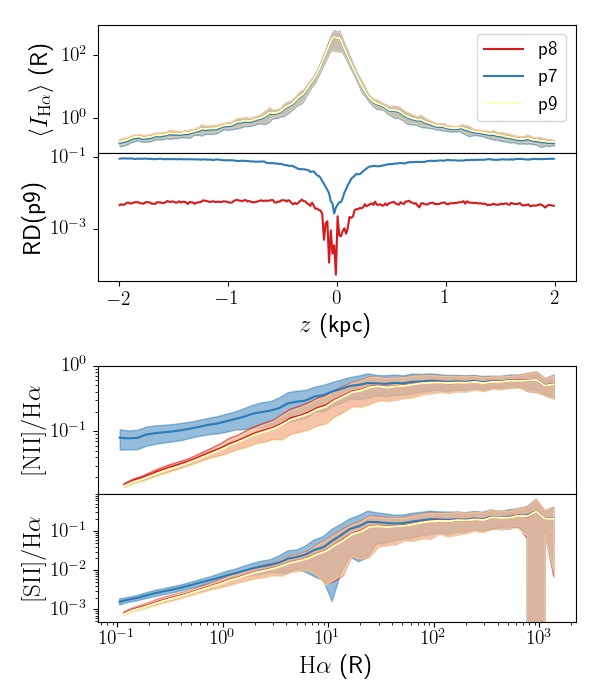}
\caption{\emph{Top~:} H$\alpha{}$ intensity profiles as a function of height and
relative difference for simulation with three different values of the photon
packet number: $10^7$ (p7), $10^8$ (p8) and $10^9$ (p9). \emph{Bottom~:} line
ratios for the same simulations.
\label{fig_emission_lines_convergence}}
\end{figure}

\begin{figure*}
\centering{}
\includegraphics[width=\textwidth]{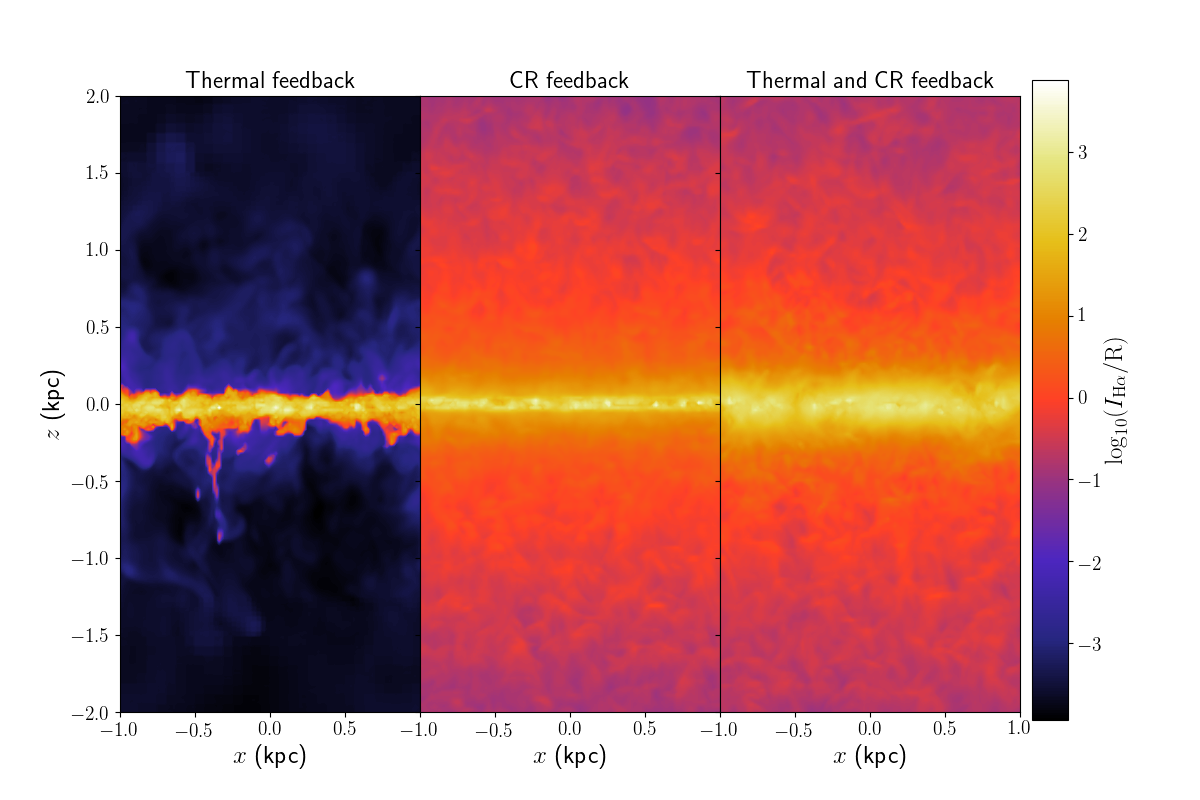}
\caption{H$\alpha{}$ intensity maps for the three models at $t = 250$~Myr. The
H$\alpha{}$ intensity is given in Rayleighs (R); 1~R corresponds to a column
emissivity of $10^6~{\rm{}photons}~{\rm{}cm}^{-2}~{\rm{}s}^{-1}~{\rm{}sr}^{-1}$.
\label{fig_emission_lines_Halpha}}
\end{figure*}

\begin{figure}
\centering{}
\includegraphics[width=0.5\textwidth]{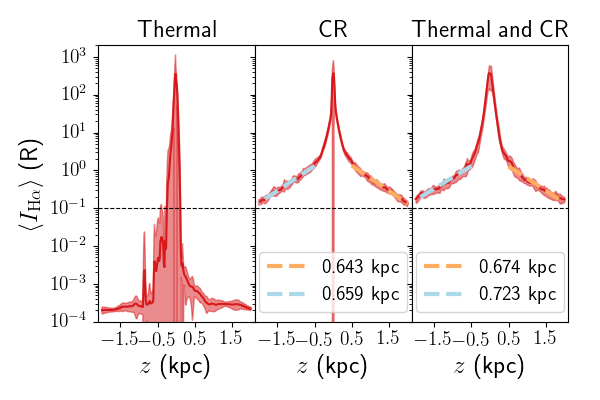}
\caption{H$\alpha{}$ intensity profiles as a function of height for the three
feedback models. The shaded regions show the scatter within planar regions of
equal scale height. The dashed lines show exponential fits to the part of the
curve with $|z| > 500$~pc (we fitted the parts below and above the disc
separately). The scale heights derived from these fits are indicated in the
legend. The black dashed line indicates the sensitivity limit for the Wisconsin
H$\alpha{}$ Mapper (WHAM), a representative instrument used to observe emission
lines in the Milky Way disc. Note that the large scatter near the midplane
causes some of the shaded regions to have negative lower bounds, which can not
be represented on the logarithmic scale.
\label{fig_emission_lines_scaleheight}}
\end{figure}

To calculate emission lines, we used the full version of \textsc{CMacIonize},
which treats the temperature of the ISM as a variable rather than a constant,
and uses heating and cooling terms to obtain a converged temperature and
ionisation state for each cell in the grid. Since the abundances of most
coolants are significantly lower than those of hydrogen and helium, we increased
the number of photons for these simulations with a factor of 10 to get better
signal to noise ratios for the coolant ionic fractions. The
convergence of the resulting H$\alpha{}$ profiles and forbidden line emission
ratios is shown in \figureref{fig_emission_lines_convergence} for the model with
both thermal and cosmic ray feedback. We see that $10^8$ photon packets is
sufficient to get converged H$\alpha{}$ profiles and emission line ratios.

\figureref{fig_emission_lines_Halpha} shows simulated maps of the H$\alpha{}$
intensity for our different models. It is clear that the H$\alpha{}$ emission
is significantly more extended in the simulations including cosmic ray feedback.
Note that the H$\alpha{}$ intensities we find here are significantly higher than
those found by \citet{2014Barnes} (see Section 3.4.2 below).

\figureref{fig_emission_lines_scaleheight} shows horizontally averaged
H$\alpha{}$ profiles. It is clear that the simulation with only thermal feedback
has very low intensities at even moderate heights above the midplane, which
indicates that this model does not have an extended ionized disc. By fitting an
exponential profile of the form ${\rm{}H}\alpha{}(z) = {\rm{}H}\alpha{}_0
\exp(-|z|/z_s)$ at heights above and below $500~{\rm{}pc}$ for the models with
cosmic ray feedback, we can determine the H$\alpha{}$ scale height $z_s$ for
those models. This yields values of $0.643~{\rm{}kpc}$ and $0.659~{\rm{}kpc}$
for the model with only cosmic ray feedback, and $0.674~{\rm{}kpc}$ and
$0.723~{\rm{}kpc}$ for the model with both thermal and cosmic ray feedback.
These values are in line with observed H$\alpha{}$ scale heights in the Milky
Way, which range from $\sim{}100~{\rm{}pc}$ in the inner galaxy
\citep{2005Madsen}, over $\sim{}400~{\rm{}pc}$ in the near spiral arms
\citep{1999Haffner, 2014Hill, 2017Krishnarao} and values of
$\sim{}700~{\rm{}pc}$ in the local solar neighbourhood \citep{2008Gaensler,
2009Savage}, to values of more than $1~{\rm{}kpc}$ in the far Carina arm
\citep{2017Krishnarao}. The values we find are also significantly higher than
the $\sim{}200~{\rm{}pc}$ scale heights found in \citet{2014Barnes}.

\begin{figure}
\centering{}
\includegraphics[width=0.5\textwidth]{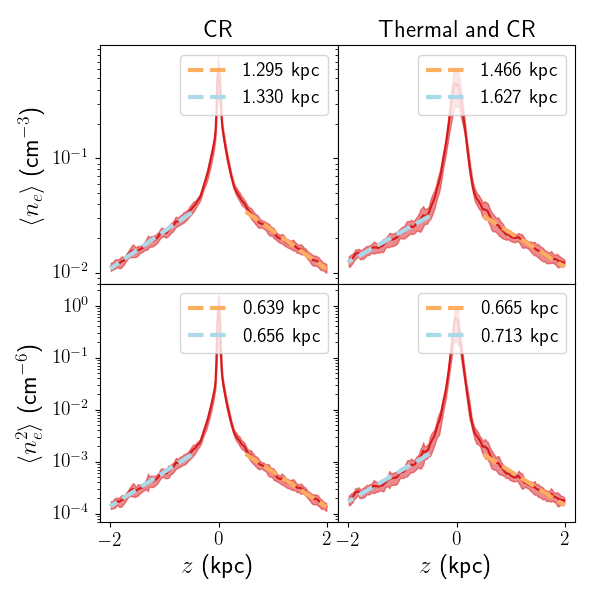}
\caption{Average electron density $n_e$ (\emph{top}) and squared electron
density $n_e^2$ (\emph{bottom}) as a function of height above the disc. The red
line shows the average value within slices of equal height, the shaded region
shows the scatter within each slice. The dashed lines indicate exponential fits
to the part of the curve with $|z| > 500$~pc (we fitted the parts below and
above the disc separately). The scale heights derived from these fits are
indicated in the legend.
\label{fig_emission_lines_Halpha_vs_ne}}
\end{figure}

\citet{2008Gaensler} propose to measure the filling factor of ionized gas by
comparing H$\alpha{}$ scale heights with electron scale heights obtained from
pulsar dispersion measurements. Assuming an ionizing gas filling factor which is
constant as a function of height, the electron scale height $z_{s,n_e}$ (which
traces the electron density $n_e$) and H$\alpha{}$ scale height $z_{s, n_e^2}$
(which scales as $n_e^2$) should be linked by $z_{s,n_e} / z_{s,n_e^2} = 2$.
However, if these scale heights are inferred from integrated quantities, then
the presence of neutral regions in the integral could skew this ratio to higher
values for low altitudes. We test this for our simulations by fitting scale
heights to $n_e(z)$ and $n_e^2(z)$, as shown in
\figureref{fig_emission_lines_Halpha_vs_ne}. As can be seen, the $z_{s,n_e^2}$
values we find are in good agreement with the integrated H$\alpha{}$ values.
Furthermore, the $z_{s,n_e}$ values we find are consistent with a ratio of 2.

\begin{figure*}
\centering{}
\includegraphics[width=\textwidth]{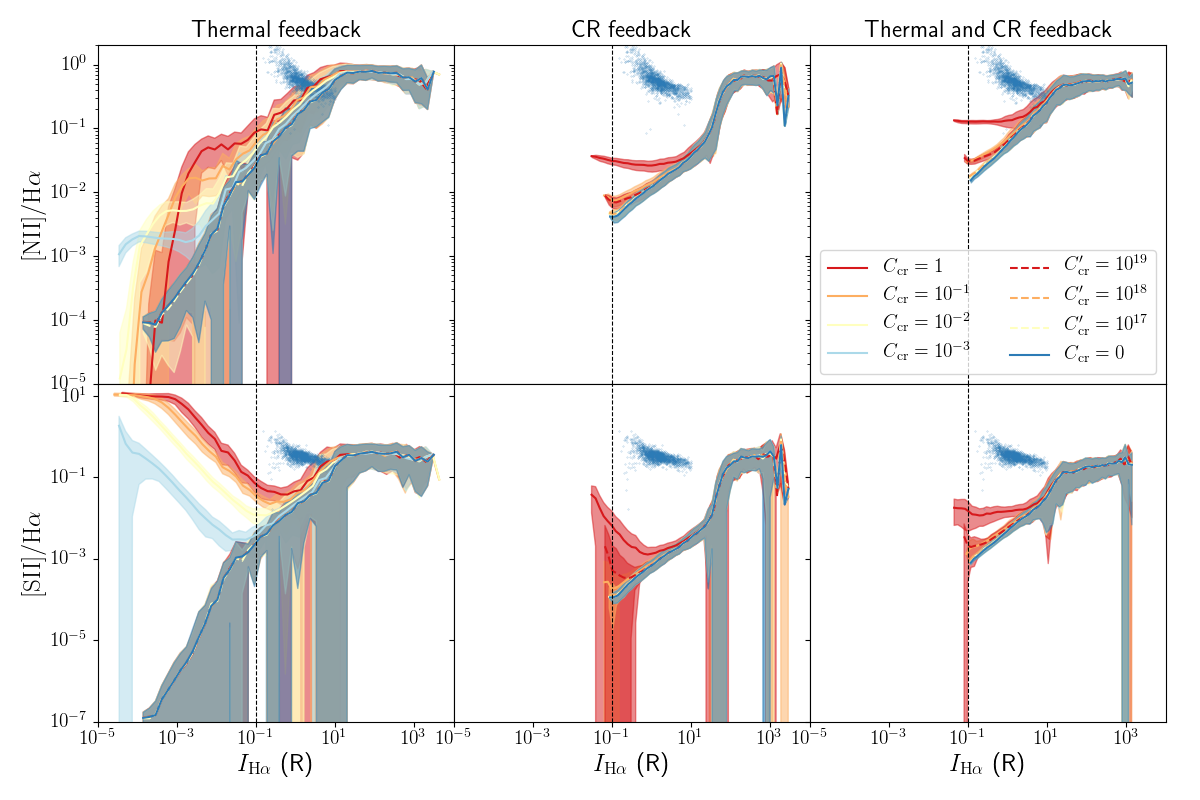}
\caption{Line ratios for the simulations with and without cosmic ray heating.
The different lines correspond to different values of the cosmic ray heating
factors $C_{\rm{}cr}$ or $C'_{\rm{}cr}$, as indicated in the legend. The blue
line corresponds to the line ratios of the reference model without cosmic ray
heating. All values have been binned in 50 logarithmically spaced bins. The
shaded regions show the scatter within the bins. The black dashed line is the
WHAM sensitivity limit, while the blue dots represent observational data from
the WHAM survey \citep{1999Haffner}. Note that the large scatter for some
H$\alpha{}$ intensities causes the shaded regions to have negative lower bounds,
which can not be represented on the logarithmic scale.
\label{fig_emission_lines_cr_line_ratios}}
\end{figure*}

\begin{figure}
\centering{}
\includegraphics[width=0.5\textwidth]{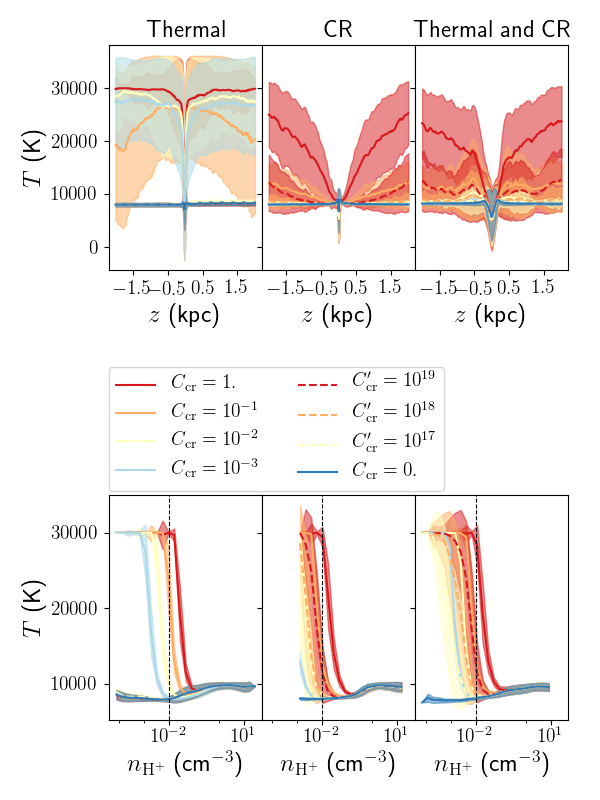}
\caption{\emph{Top~:} average temperature profiles for the simulations with and
without cosmic ray feedback. The coloured lines are the results for different
values of the cosmic ray heating factors $C_{\rm{}cr}$ or $C'_{\rm{}cr}$, as
indicated in the legend. The blue line $C_{\rm{}cr}=0$ is the reference model
without cosmic ray heating. The shaded regions show the scatter within slices of
equal scale height. \emph{Bottom~:} temperature as a function of ionised
density for the same simulations. The results have been binned in 50
logarithmically spaced bins; the shaded regions represent the scatter within the
bins. The vertical dashed lines represent the density threshold below which the
gas will likely be shock heated, an effect which we do not take into account in
our treatment.
\label{fig_emission_lines_cr_temperature}}
\end{figure}

When we look at emission line ratios, the correspondence between model and
observations is much less favourable however.
\figureref{fig_emission_lines_cr_line_ratios} shows the
$[{\rm{}NII}]/{\rm{}H}\alpha{}$ and $[{\rm{}SII}]/{\rm{}H}\alpha{}$ ratios as a
function of the H$\alpha{}$ intensity. We notice a clear upwards trend for both
line ratios for increasing H$\alpha{}$ intensity. This trend is completely
opposite to the downwards trend found in observations \citep{1999Haffner}, and
agrees with earlier results found by \citet{2014Barnes}. Since
$[{\rm{}NII}]/{\rm{}H}\alpha{}$ and $[{\rm{}SII}]/{\rm{}H}\alpha{}$ trace warm
ionised gas, the lack of high line ratios for low H$\alpha{}$ intensities
indicates that the temperature at high altitudes above the disc (where the
H$\alpha{}$ intensity is low) is too low in our models. This can also be seen in
\figureref{fig_emission_lines_cr_temperature}, where the equilibrium gas
temperature is shown as a function of altitude. The equilibrium temperature in
our models only increases very slowly across the ionised disc. If we want our
models to line up with the observations, we hence need an extra heating
mechanism that can increase the temperature at high altitudes.

\subsubsection{Cosmic ray heating}

One possible source of heating would be heating by cosmic rays. The cosmic ray
model employed by the SILCC simulations introduces an extra pressure term and
associated adiabatic heating and cooling due to cosmic rays, but does not
constitute a direct heating term \citep{2016Girichidisb}. This means we cannot
directly use the simulation snapshots to provide us with the required additional
heating. We know however that cosmic rays can directly transfer energy to the
gas by the generation of dampened Alfv\'{e}n waves \citep{1971Wentzel}, a
process which scales with $n_e^{-1/2}$, where $n_e$ is the electron density in
the gas \citep{2013Wiener}.

To test if cosmic ray heating can explain the observed line ratios, we run
additional photoionisation simulations with the advanced model that treats both
the ionisation balance and the temperature as a variable, and include an extra
heating term that scales with $n_e^{-1/2}$:
\begin{equation}
H_{\rm{}cr} = \left(1.2 \times{} 10^{-27} {\rm{}erg}~{\rm{}cm}^{-9/2}
{\rm{}s}^{-1} \right) C_{\rm{}cr} n_e^{-1/2} {\rm{}e}^{-|z|/h_{\rm{}cr}},
\end{equation}
where $C_{\rm{}cr}$ is a constant parameter of our model, $z$ is the height
above the disc, $h_{\rm{}cr}$ is the scale height of the cosmic ray heating, and
the numerical factor was based on the factor given in \citet{2013Wiener}. As in
\citet{2013Wiener}, we choose a value $h_{\rm{}cr}=1.333$~kpc. We will consider
different values of the parameter $C_{\rm{}cr}$.

We also implemented a second cosmic ray heating scheme that is based
on the more exact expression for the cosmic ray heating\citep{2013Wiener}:
\begin{equation}
H_{\rm{}cr} = \boldsymbol{v_A} . \boldsymbol{\nabla{}}P_{\rm{}cr},
\end{equation}
where $\boldsymbol{v_A}$ is the Alfv\'{e}n speed ($\boldsymbol{v_A} =
\frac{\boldsymbol{B}}{\sqrt{\mu{}_0 m_p n_e}}$, with $\boldsymbol{B}$ the
magnetic field, $\mu{}_0$ the vacuum permeability and $m_p$ the proton mass),
and $P_{\rm{}cr}$ is the cosmic ray pressure ($P_{\rm{}cr} = (\gamma{}_{\rm{}cr}
- 1) e_{\rm{}cr}$, with $\gamma{}_{\rm{}cr} = 4/3$ the cosmic ray polytropic
index, and $e_{\rm{}cr}$ the cosmic ray energy density). $\boldsymbol{B}$ and
$e_{\rm{}cr}$ are present in the SILCC outputs, so our alternative heating term
then can be formulated as
\begin{equation}
H_{\rm{}cr} = \left(1.2 \times{} 10^{-27} {\rm{}erg}~{\rm{}cm}^{-9/2}
{\rm{}s}^{-1} \right) C'_{\rm{}cr} \frac{\boldsymbol{B} .
\boldsymbol{\nabla{}}e_{\rm{}cr}}{\sqrt{n_e}},
\end{equation}
with $C'_{\rm{}cr}$ a constant parameter. This heating term will be trivially
zero for the SILCC model without cosmic ray feedback.

Since the heating terms described above diverge for very low values of the
electron density, we have to be careful when adding them to our temperature
calculation, as low electron densities do not only occur in ionised cells with
low densities, but also in denser, neutral cells, where the electron density is
low due to the high neutral fraction of the gas. We therefore, before entering
the complex temperature iteration for a cell, compute the neutral fraction at a
fixed temperature of 8000~K for that cell. If the cell has a neutral fraction
that is higher than a cut off value of 0.75, we assume the cell is fully neutral
and set its temperature to 500~K. In this case we do not enter the detailed
cooling and heating routine, and avoid problems with finding the neutral
fraction in cells with high densities and low electron densities.

Similarly, we also need to take care when heating the cells to very high
temperatures, as we only included the relevant heating and cooling processes up
to temperatures of about 30,000~K. We therefore manually reset the temperature
to 30,000~K if cosmic ray heating pushes it to higher values.

\begin{figure}
\centering{}
\includegraphics[width=0.5\textwidth]{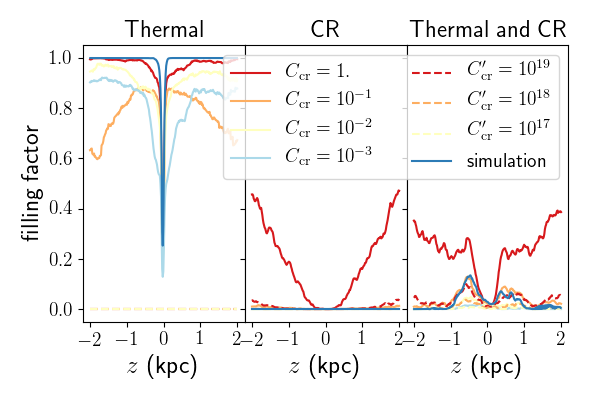}
\caption{Hot gas filling factors for the simulations with cosmic ray heating.
The different lines correspond to different values of the cosmic ray heating
factor $C_{\rm{}cr}$, as indicated in the legend. The blue line corresponds to
the hot gas filling factor found in the SILCC snapshots themselves.
\label{fig_emission_lines_cr_filling_factor}}
\end{figure}

\figureref{fig_emission_lines_cr_temperature} shows the temperature profiles for
various strengths of the cosmic ray heating. It is clear that cosmic ray heating
is able to push the average temperature to higher values. However, it is also
worth noting that a significant fraction of the gas ends up at 30,000~K, the
imposed temperature limit, especially for cells at higher altitudes above the
disc. These cells will have very low H$\alpha{}$ intensities, so they do not
contribute significantly to the observed line ratios.
\figureref{fig_emission_lines_cr_filling_factor} shows the hot gas filling
factor in each horizontal slice of the simulation, defined as the fraction of
the $128\times{}128$ cells in the slice with a temperature of 30,000~K. For
comparison, we also show the hot gas filling factor in the actual SILCC
snapshots. It is clear that the models with high cosmic ray feedback contain a
lot of cells with very high temperatures. To reproduce the hot gas filling
factor of the SILCC simulations in the models with cosmic ray feedback, we would
need a cosmic ray heating parameter with about 10\% of the value advocated in
\citet{2013Wiener}, or the more advanced cosmic ray heating model
with the highest parameter value.

Note that both cosmic ray heating models produce similar results. As
can be seen from the bottom panel of
\figureref{fig_emission_lines_cr_temperature}, both mechanisms predominantly
heat low density ionised gas, and cause a sharp increase in temperature at a
density threshold set by the value of the heating parameter.

Also note that gas with densities lower than $\sim{}10^{-2}~{\rm{}cm}^{-3}$ will
most likely be shock heated and ionised to temperatures $>10^5~{\rm{}K}$, an
effect that we did not incorporate in our post-processing treatment. Therefore,
we find that all but the strongest cosmic ray heating realisations only heat
gas that would already be heated by other physical mechanisms.

Finally, \figureref{fig_emission_lines_cr_line_ratios} shows the line ratios for
the different cosmic ray heating models. It is clear that the cosmic ray heating
does not affect the high H$\alpha{}$ intensity end of the curve, which
corresponds to radiation emitted by the central disc. Cosmic ray heating does
affect the low intensity line ratios in the diffuse ionised disc, and
effectively changes the sign of the slope of the curve, leading to a trend that
is more in line with the observed trend, although the effect is only noticeable
for high values of the cosmic ray heating factor. This seems to
indicate that the cosmic ray heating does not affect the temperature of the
intermediate density ionised gas which is responsible for the observed
H$\alpha{}$ emission.

\subsubsection{Lower total luminosity}

\begin{figure*}
\centering{}
\includegraphics[width=\textwidth]{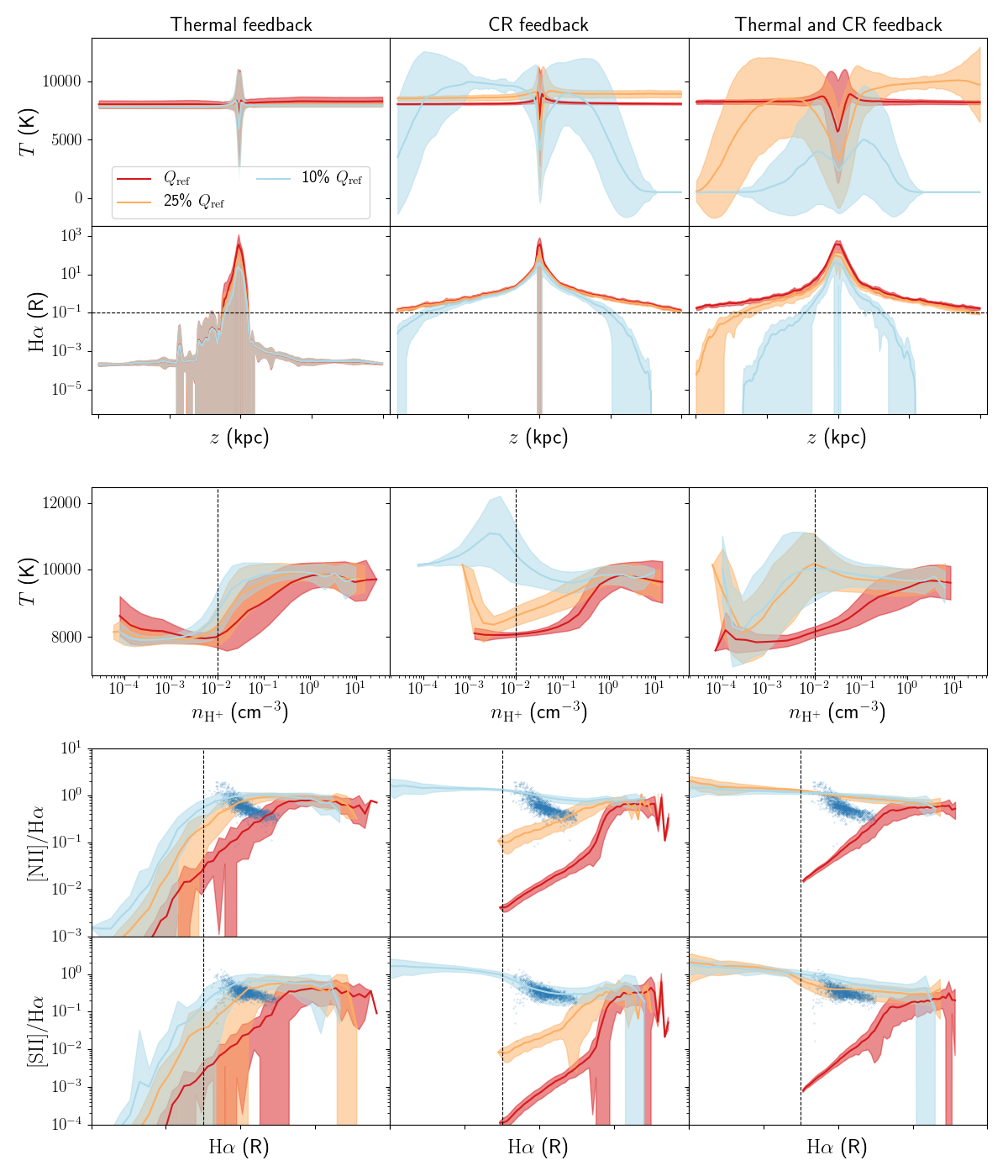}
\caption{Average temperature and ${\rm{}H}\alpha{}$ intensity as a function of
height above the disc, temperature as a function of ionised density, and
$[{\rm{}NII}]/{\rm{}H}\alpha{}$ and $[{\rm{}SII}]/{\rm{}H}\alpha{}$ line
strenghts as a function of ${\rm{}H}\alpha{}$ intensity for the models with
different values for the total ionising luminosity
($Q_{\rm{}ref}$ represents
the reference model with an ionising luminosity of
$4.26\times{}10^{49}~{\rm{}s}^{-1}$ per ionising source).
The full lines represent the
binned average values as indicated on the legend, the shaded regions represent
the scatter within the bins. The dashed lines in the middle row represent the
density threshold below which the gas will likely be shock heated, an effect
we do not take into account in our treatment. The dashed lines in the bottom two
rows represent the sensitivity limit of
the WHAM instrument, while the blue dots represent observational data from the
WHAM survey \citep{1999Haffner}. Note that the large scatter for some scale
heights and H$\alpha{}$ intensities causes the shaded regions to have negative
lower bounds, which can not be represented on the logarithmic scale.
\label{fig_emission_lines_low_luminosity}}
\end{figure*}

We require a relatively high luminosity to ionise out to large scale heights,
compared to the values that were used by \citet{2014Barnes}. This leads to
H$\alpha{}$ luminosities that are high compared to observed H$\alpha{}$
luminosities. It is instructive to see how the temperature and line ratios
change if a lower luminosity is used.

To this end, we run a set of simulations with the full version of the code,
using lower luminosities for the ionising sources: a version with an ionizing
luminosity of $4.26 \times{} 10^{48}~{\rm{}s}^{-1}$ per ionising source (10\% of
the generic value), and a version with a luminosity of $1.065 \times{}
10^{49}~{\rm{}s}^{-1}$ per ionising source (25\% of the generic value).

\begin{table}
\caption{Scale height of the H$\alpha{}$ disc for the simulations with lower
total ionising luminosities.
\label{table_low_luminosity_scale}}
\begin{tabular}{c c c}
\hline
ionising luminosity & scale height $z<0$ & scale height $z>0$ \\
(${\rm{}s}^{-1}~{\rm{}source}^{-1}$) & (kpc) & (kpc) \\
\hline
TH1\_CR0 \\
\hline
$4.26 \times{} 10^{49}$ & $0.931$ & $2.064$ \\
$1.065 \times{} 10^{49}$ & $0.964$ & $2.065$ \\
$4.26 \times{} 10^{48}$ & $0.980$ & $2.077$ \\
\hline
TH0\_CR1 \\
\hline
$4.26 \times{} 10^{49}$ & $0.659$ & $0.643$ \\
$1.065 \times{} 10^{49}$ & $0.662$ & $0.641$ \\
$4.26 \times{} 10^{48}$ & $0.352$ & $0.051$ \\
\hline
TH1\_CR1 \\
\hline
$4.26 \times{} 10^{49}$ & $0.723$ & $0.674$ \\
$1.065 \times{} 10^{49}$ & $0.204$ & $0.585$ \\
$4.26 \times{} 10^{48}$ & $0.055$ & $0.056$ \\
\hline
\end{tabular}
\end{table}

The results of these simulations are shown in
\figureref{fig_emission_lines_low_luminosity}. As expected, the lower
luminosities lead to more noisy temperature and H$\alpha{}$ profiles, as the
sources are no longer able to ionise the entire low density disc. However, for
the SILCC models including cosmic ray feedback, the H$\alpha{}$ scale height for
the simulation with 25\% of the total ionizing luminosity is still comparable to
the full luminosity result (as can be seen from
\tableref{table_low_luminosity_scale}), and hence in line with observed scale
heights in the Milky Way. Furthermore, the runs with lower intensity
show a clear increase in temperature for intermediate density ionised gas,
which in turns translates into line ratios that increase with decreasing
H$\alpha{}$ intensity.

We hence find that a decrease in total ionising luminosity can actually lead to
higher local temperatures and can hence explain the observed line ratios.
This counter-intuitive result can be explained by the hardening of
the ionising spectrum for large scale heights \citep{2004Woodb}~: low frequency
photons are preferentially absorbed at low heights, so that the spectrum that
reaches higher heights contains relatively more high frequency hydrogen
ionising photons, which heat the gas to higher temperatures. If the total
ionising luminosity is high, the fraction of photons absorbed at low heights
will be low, and the hardening of the spectrum will happen over a large scale.
However, if the luminosity is just enough to ionise out to the boundary of the
simulation box, we see the full hardening of the spectrum and resulting rise in
temperature, while still keeping the gas ionised. There is however some fine
tuning required: decreasing the total luminosity further to 10\% of the generic
value significantly decreases the H$\alpha{}$ scale length, and leads to results
that are no longer in line with observations.

\section{Conclusion}

In this work, we post-processed star forming disc simulations of the SILCC
project with the new Monte Carlo radiative transfer code \textsc{CMacIonize}. We
showed that the more extended galactic disc present in MHD simulations that
include cosmic ray feedback naturally leads to larger scale heights for the
H$\alpha{}$ emission from the DIG in these discs, provided that enough ionising
radiation is present to ionise out to large scale heights. This more extended
disc is a direct consequence of the specific way cosmic rays couple to the
hydrodynamics of the ISM, which makes it possible to more efficiently heat the
gas in a larger region compared to thermal feedback.

However, most of our simulations are unable to reproduce observed nitrogen and
sulphur line ratios, even when additional heating due to cosmic rays is included
in the model, since these models only affect the temperature of low
density ionised gas that does not contribute to observed line ratios. Only if
we use a total ionising luminosity that is marginally sufficient to ionise the
DIG, we are able to increase the temperature of intermediate
density ionised gas and reproduce observed Milky Way line ratios. This fine
tuning can have several explanations:
\begin{itemize}
  \item{} The ionising radiation itself is partially responsible for setting the
  scale height of the ionised gas. This would naturally lead to a correlation
  between the total ionising luminosity and the scale height. However, we cannot
  study this effect in this work, as there is no direct dynamical link between
  the SILCC simulations and our post-processing tool. We plan to repeat our
  analysis for the \citet{2017Peters} simulations that do include
  photoionization feedback.
  \item{} There is another physical heating mechanism that is responsible for
  the extra heating that is necessary to explain observed line ratios,
  and that affects intermediate rather than low density ionised
  gas. In   this case we would not need to fine tune the ionising luminosity.
  \item{} The dense gas surrounding young O stars aborbs about 75\% of the
  ionising radiation from the star, so that only 25\% is left to ionise the
  more diffuse surrounding gas. As we do not resolve the densest gas, this is an
  effect that is likely. But to find more accurate values of the ionising escape
  fraction, more detailed simulations of star forming clouds are necessary.
\end{itemize}

We hence have no satisfactory explanation for the observed line ratios in the
DIG of the Milky Way, and leave this for future work. We do find that the
observed decreasing line ratios of nitrogen and sulphur with increasing
H$\alpha{}$ intensity are not necessarily explained by an increasing average ISM
temperature with increasing scale height, as is often assumed. Instead, these
line ratios could also be consistent with an increase in temperature dispersion
with increasing scale height, with an overall constant or decreasing average
temperature.

\section*{Acknowledgements}
We want to thank the anonymous referee for constructive and insightful remarks
that significantly improved the quality of this work.
BV and KW acknowledge support from STFC grant ST/M001296/1. PG acknowledges
support from the DFG Priority Program 1573 Physics of the Interstellar Medium as
well as funding from the European Research Council under ERC-CoG grant
CRAGSMAN-646955.

\bsp    
\label{lastpage}

\end{document}